\documentclass[prl,aps,floats,twocolumn,showpacs,nofootinbib]{revtex4}

\usepackage{amssymb}
\usepackage{amsmath}

\usepackage{graphicx}
\usepackage{graphics}
\usepackage{dcolumn}
\usepackage{color}
\usepackage{rotate}
\usepackage{fancyhdr} 
\usepackage{hyperref}
\usepackage{indentfirst}

\usepackage{umoline}
\usepackage{ulem}

\def\be{\begin{equation}}
\def\ee{\end{equation}}
\def\bea{\begin{eqnarray}}
\def\eea{\end{eqnarray}}
\def\ba{\begin{eqnarray}}
\def\ea{\end{eqnarray}}
\def\nn{\nonumber}

\newcommand{\Mpl}{M_{\rm pl}}

\definecolor{darkred}{rgb}{.743,0,0}

\newcommand{\refeq}[1]{Eq.~(\ref{eq:#1})}          
          
\newcommand{\reffig}[1]{Fig.~\ref{fig:#1}}

\begin{document}

\title{Reheating constraints to inflationary models}

\author{Liang Dai, Marc Kamionkowski, and Junpu Wang}
\affiliation{Department of Physics and Astronomy, 
Johns Hopkins University, 3400 N.\ Charles St., Baltimore, MD 21218
}

\date{\today}

\begin{abstract}
Evidence from the BICEP2 experiment for a significant
gravitational-wave background has focussed attention on 
inflaton potentials $V(\phi) \propto \phi^\alpha$ with
$\alpha=2$ (``chaotic'' or ``$m^2\phi^2$'' inflation) or with
smaller values of $\alpha$, as may arise in axion-monodromy models.
Here we show that reheating considerations may
provide additional constraints to these models.  The
reheating phase preceding the radiation era is modeled by an effective
equation-of-state parameter $w_{\rm re}$.  The canonical
reheating scenario is then described by $w_{\rm re}=0$.   The simplest
$\alpha=2$ models are consistent with $w_{\rm re} =  
0$ for values of $n_s$ well within the current $1\sigma$ range.
Models with $\alpha=1$ or $\alpha=2/3$ require a more exotic 
reheating phase, with $-1/3<w_{\rm re}<0$, unless $n_s$
falls above the current $1\sigma$ range.  Likewise,
models with $\alpha=4$
require a physically implausible $w_{\rm re}>1/3$, unless $n_s$
is close to the lower limit of the $2\sigma$ range.
For $m^2\phi^2$ inflation and canonical reheating as a benchmark,
 we derive a relation
$\log_{10}\left(T_{\rm re}/10^6\,{\rm GeV} \right) \simeq
2000\,(n_s-0.96)$ between the reheat temperature $T_{\rm re}$
and the scalar spectral index $n_s$.  Thus, if $n_s$ is close to
its central value, then $T_{\rm re}\lesssim 10^6$~GeV, just above
the electroweak scale.  If the reheat temperature is
higher, as many theorists may prefer, then the 
scalar spectral index should be closer to
$n_s\simeq0.965$ (at the pivot scale $k=0.05\,{\rm
Mpc}^{-1}$), near the upper limit of the $1\sigma$ error
range.  Improved precision in the measurement of $n_s$
should allow $m^2\phi^2$, axion-monodromy, and $\phi^4$ models
to be distinguished, even without precise measurement of $r$,
and to test the $m^2\phi^2$ expectation of $n_s\simeq0.965$.
\end{abstract}

\maketitle

\noindent
{\it Introduction.}\, The imprint of inflationary gravitational
waves in the cosmic microwave background polarization
\cite{Kamionkowski:1996ks} reported by the BICEP2 collaboration
\cite{Ade:2014xna} implies, if confirmed, that the inflaton field $\phi$
traversed a distance large compared with the Planck mass during inflation
\cite{Turner:1993su,Lyth:1996im}.  One particularly simple and
elegant model for large-field inflation is ``$m^2\phi^2$'' inflation
\cite{Piran,Belinsky:1985zd} (derived originally as a simple
example of chaotic inflation \cite{Linde:1983gd}), in which the
inflaton potential is simply a quadratic function of $\phi$.
Ref.~\cite{Creminelli:2014oaa} recently argued that this is
perhaps the simplest and most elegant model.  They then
derived a consistency relation between the scalar spectral index
(now constrained to be $n_s-1=-0.0397\pm
0.0073$~\cite{Ade:2013uln}) and tensor-to-scalar ratio (roughly
$r\sim0.2$ according to Ref.~\cite{Ade:2014xna}) that can be
tested with higher-precision measurements of $n_s$ and in
particular of $r$. Another promising candidate large-field
model, axion monodromy which suggests a
potential $V\propto \phi$ \cite{McAllister:2008hb} or $V\propto
\phi^{2/3}$ \cite{Silverstein:2008sg}, has also been receiving
considerable attention.  We parametrize all these models by a
power-law potential $V\propto \phi^\alpha$.

Here we point out that consideration of the process by which the
Universe reheats may provide additional constraints to these
models~\cite{Dodelson:2003vq,Martin:2010kz,Adshead:2010mc,Mielczarek:2010ag,Easther:2011yq}. 
After inflation ends, there must be a period of
reheating (see Ref.~\cite{Allahverdi:2010xz} for a review) when
the the energy stored in the inflaton field is converted to a
plasma of relativistic particles after which the standard
radiation-dominated evolution of the early Universe takes over.
Although the physics of reheating is highly uncertain and
unconstrained, there is a simple canonical scenario
\cite{elementary-reheating} whereby
the cold gas of inflaton particles that arise from coherent oscillation
of the inflaton field about the minimum of a quadratic potential decay to
relativistic particles. This scenario implies a reheating era
that lasts for a time $\sim \Gamma^{-1}$, where $\Gamma$ is the
inflaton-decay rate, and in which the effective
equation-of-state parameter (in which the energy density scales
with scale factor $a$ as $\rho \propto a^{-3(1+w_{\rm re})}$) is
$w_{\rm re}=0$.  The radiation-dominated era is then initiated
at a temperature $T_{\rm re} \sim (\Gamma\Mpl)^{1/2}$.  Still,
there are more complicated possibilities.  For example,
resonant~\cite{Kofman:1994rk,Kofman:1997yn} or
tachyonic~\cite{Greene:1997ge} instabilities can lead to a short
preheating phase of rapid and violent dissipations by exciting
inhomogeneous modes.  After preheating, inhomogeneous modes of
the inflaton or its decay products could become
turbulent~\cite{turbulent-thermalization} and eventually evolve
to a state of equilibrium. Numerical studies of this
thermalization phase suggest a range of variation $0 \lesssim
w_{\rm re}\lesssim 0.25$~\cite{Podolsky:2005bw}. The bottom
line, though, is that $w_{\rm re}>-1/3$ is needed to end inflation,
but $w_{\rm re}>1/3$ is difficult to conceive since it requires
a potential dominated by high-dimension operators (higher than
$\phi^6$) near its minimum, unnatural from a
quantum-field-theoretical point of view.

In this Letter, we show that current measurements of $n_s$ seem
to favor $m^2\phi^2$ inflation over axion-monodromy inflation.
If $n_s$ is within its current $1\sigma$ error range,
then axion-monodromy models require an extended phase of
reheating involving exotic physics with $w_{\rm re}<0$.  Axion
monodromy is consistent with canonical reheating only if $n_s$
is above the current $1\sigma$ range.
Moreover, if $m^2\phi^2$ inflation occurred and was followed by
canonical reheating, then $n_s=0.96$ (its central value) implies
a reheat temperature just above the electroweak scale.  If
the reheat temperature was considerably higher, as may be
required to accommodate models that explain the baryon
asymmetry, then $m^2\phi^2$ inflation (with a high reheat temperature)
predicts a value $n_s\simeq0.965$, at the high end of the
currently allowed $1\sigma$ range, and a prediction that may be testable
with future CMB data and galaxy surveys.
As we will see below, these conclusions are robust to the current 
order-unity uncertainty in $r$.

\begin{figure}[ht!]
\centering
\hspace{-0.65cm}
\includegraphics[scale=0.4]{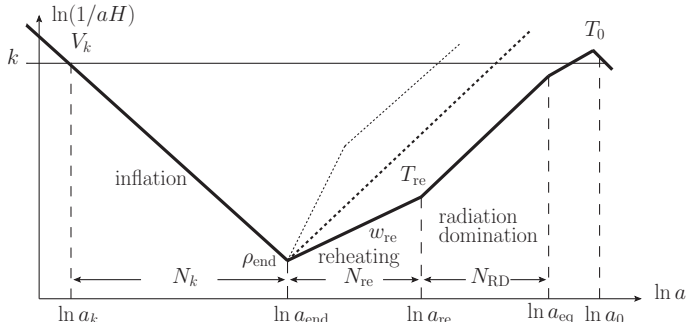}
\caption{The evolution of the comoving Hubble scale $1/aH$. The
     reheating phase connects the inflationary phase and the
     radiation era. Compared to instantaneous reheating (thick
     dotted curve), a reheating equation-of-state parameter $w_{\rm
     re}<1/3$ implies more post-inflationary $e$-folds of
     expansion. Fewer post-inflationary $e$-folds requires
     $w_{\rm re}>1/3$ (thin dotted curve).}
\label{fig:comoving-scale}
\end{figure}

We start by sketching the cosmic expansion history in
\reffig{comoving-scale}. At early times, the inflaton field
$\phi$ drives the quasi--de-Sitter phase for $N_k$ $e$-folds of
expansion. The comoving horizon scale decreases as $\sim a^{-1}$.
The reheating phase begins once the accelerated
expansion comes to an end and the comoving horizon starts to
increase. After another $N_{\rm re}$ $e$-folds of expansion,
the energy in the inflaton field has been completely
dissipated into a hot plasma with a reheating temperature
$T_{\rm re}$. Beyond that point, the Universe expands under 
radiation domination for another $N_{\rm RD}$ $e$-folds, before
it finally makes a transition to matter domination.

It is clear from \reffig{comoving-scale} that the number of
$e$-folds between the time that the current comoving horizon
scale exited the horizon during inflation and the end of
inflation must be related to the number of $e$-folds between the
end of inflation and today if the dependence of $(aH)^{-1}$ on
$a$ during reheating is known.  The expansion history also allows
us to trace the dilution of the energy density in the
Universe. To match the energy density during inflation, as fixed
by $r$, to the energy density today, a second 
relation must be satisfied. These two matching conditions, for
scale and for energy density, respectively, underly the
arguments that follow.

\noindent
{\it Quantitative analysis.}\, We consider power-law potentials
\bea
     V(\phi) = \frac12 m^{4-\alpha} \phi^\alpha,
\eea
for the inflaton, with power-law index $\alpha$ and mass parameter $m$.
From the attractor evolution of the inflaton field $3H\dot\phi +
V_{,\phi} \simeq 0$, one can determine the number
\bea
     N = \int^{\phi_{\rm end}}_{\phi} \frac{H  d\phi}{\dot\phi}
     \simeq \frac{\phi^2 - \phi^2_{\rm end}}{2\alpha \Mpl^2} \simeq
     \frac{\phi^2}{2\alpha \Mpl^2},
\eea 
of $e$-folds from the time that the field value is $\phi$ until
the end of inflation.  Note that the field value at the end of
inflation $\phi_{\rm end}$ is small compared to that during
slow-roll. The conventional slow-roll parameters are then given
by 
\bea
     \epsilon = \alpha/(4N),\quad {\rm and}\quad \eta =
     (\alpha-1)/(2N).
\eea
For power-law potentials, the scalar spectral tilt $n_s-1$ and
the tensor-to-scalar ratio $r$ are inversely proportional to the
number of $e$-folds,
\bea
n_s - 1 = -(2+\alpha)/(2N),\quad r = 4\alpha/N.
\eea
Simultaneous measurements of $n_s-1$ and $r$ with high precision
in principle pin down both $N$ and $\alpha$. However, given the current
uncertainty in $r$, we treat $\alpha$ as a model input and use
$n_s-1$ to infer {\it both} $N$ and $r$. As we shall see, the precise 
value of $r$ does not affect our results.

In cosmology we observe perturbation modes on scales that are
comparable to that of the horizon. For example, the pivot
scale at which Planck determines $n_s$ lies at $k=0.05~{\rm
Mpc}^{-1}$. The comoving Hubble scale $a_k H_k = k$ when this
mode exited the horizon can be related to that of the present
time,
\bea
     \frac{k}{a_0 H_0} = \frac{a_k}{a_{\rm end}} \frac{a_{\rm
     end}}{a_{\rm re}} \frac{a_{\rm re}}{a_{\rm eq}}
     \frac{a_{\rm eq} H_{\rm eq}}{a_0 H_0} \frac{H_k}{H_{\rm
     eq}}.
\eea
Here quantities with subscript $k$ are evaluated at the time
of horizon exit. Similar subscripts refer to other epochs,
including the end of inflation (${\rm end}$), reheating
(${\rm re}$), radiaton-matter equality (${\rm eq}$) and
the present time ($0$). Using $e^{N_k}=a_{\rm end}/a_k$,
$e^{N_{\rm re}}=a_{\rm re}/a_{\rm end}$, and $e^{N_{\rm
RD}}=a_{\rm eq}/a_{\rm re}$, we obtain a constraint on the total
amount of expansion~\cite{Liddle:2003as},
\bea
\label{eq:efolds-eq-1}
   \ln\frac{k}{a_0 H_0} = - N_k - N_{\rm re} - N_{\rm RD} +
   \ln\frac{a_{\rm eq} H_{\rm eq}}{a_0 H_0} + \ln
   \frac{H_k}{H_{\rm eq}}.
\eea
The Hubble parameter during inflation is given by $H_k = \pi
\Mpl \left(r A_s\right)^{1/2}/\sqrt 2$, with the primordial
scalar amplitude $\ln(10^{10} A_s) = 3.089^{+0.024}_{-0.027}$
from Planck~\cite{Ade:2013uln}. For a given power-law index
$\alpha$, $N_k$ and $r$ are determined from $n_s-1$, and 
hence $\ln H_k$ is known.

In addition to \refeq{efolds-eq-1}, a second relation between
the various $e$-folds of expansion can be derived by tracking the
post-inflationary evolution of the energy density and
temperature. The inflaton field at the end of inflation has a
value $\phi_{\rm end}=(\alpha^2 \Mpl^2/2\epsilon_0)^{1/2}$ under
the estimate that inflation terminates at
$\epsilon=\epsilon_0 \simeq 1$, while its value during
inflation satisfies $N_k=\phi^2_k/(2\alpha\Mpl^2)$. Therefore,
the final stage of inflation phase has potential energy $V_{\rm
end}=V_k (\phi_{\rm end}/\phi_k)^\alpha$, where $V_k=3 \Mpl^2
H^2_k = (3\pi^2/2)\Mpl^4\,r A_s$. The energy density is
$\rho_{\rm end}=(1+\lambda)V_{\rm end}$, with the ratio
$\lambda=(3/\epsilon_0-1)^{-1}$ of kinetic energy to
potential energy.

The duration,
\bea
\label{eq:reheating-efoldsone}
     N_{\rm re} = [3(1+w_{\rm re})]^{-1} \ln \left( \rho_{\rm
     end} / \rho_{\rm re} \right),
\eea
of reheating determines the dilution of the energy density.
Here for simplicity we assume $w_{\rm re}$ is a constant.
The final energy density determines the reheating temperature
through $\rho_{\rm re}=(\pi^2/30) g_{\rm re} T^4_{\rm re}$, with
$g_{\rm re}$ being the effective number of relativistic species
upon thermalization. The subsequent expansion is mainly driven
by hot radiation, except for very recently non-relativistic
matter and dark energy. Although it remains a possiblity before
BBN at $z > 10^9$, for simplicity we assume that no immense
entropy production take place after $T_{\rm re}$. Under this
assumption, the reheating entropy is preserved in the CMB and
neutrino background today, which leads to the relation,
\bea
\label{eq:reheating-present}
     g_{s,\rm re} T^3_{\rm re} = \left(\frac{a_0}{a_{\rm
     re}}\right)^3 \left( 2\, T^3_0 + 6\cdot \frac78 T^3_{\nu 0}
     \right),
\eea
with the present CMB temperature $T_0=2.725~\rm K$, the
neutrino temperature $T_{\nu 0}=(4/11)^{1/3} T_0$, and the
effective number of light species for entropy $g_{s,\rm re}$ at
reheating. We therefore relate the reheating temperature to the
present CMB temperature through,
\bea
     \frac{T_{\rm re}}{T_0} = \left( \frac{43}{11 g_{s,\rm re}}
     \right)^{1/3} \frac{a_0}{a_{\rm eq}} \frac{a_{\rm
     eq}}{a_{\rm re}}.
\label{eq:reheating-presenttwo}
\eea
Combining \refeq{reheating-efoldsone}, \refeq{reheating-presenttwo},
and other relations lead to a second equation relating the
various $e$-folds,
\bea
\label{eq:efolds-eq-2}
     \frac{3(1+w_{\rm re})}{4} N_{\rm re} & = & \frac14 \ln
     \frac{30}{g_{\rm re} \pi^2} + \frac14 \ln \frac{\rho_{\rm
     end}}{T_0^4} + \frac13 \ln \frac{11 g_{s,\rm re}}{43}
     \nn\\ 
     && + \ln \frac{a_{\rm eq}}{a_0} - N_{\rm RD}.
\eea 
We now combine \refeq{efolds-eq-1} and \refeq{efolds-eq-2} and 
\bea
\label{eq:reheating-efolds}
     && N_{\rm re} = \frac{4}{1-3 w_{\rm re}} \left[ - N_k -
     \ln\frac{k}{a_0 T_0} - \frac14 \ln
     \frac{30}{g_{\rm re} \pi^2} \right. \nn\\
     && \left. - \frac13 \ln \frac{11 g_{s,\rm re}}{43} +
     \frac14 \ln\frac{\pi^2 r A_s}{6} - \frac{\alpha}{8}
     \ln\frac{r}{16 \epsilon_0} - \frac{\ln(1+\lambda)}{4} \right]. \nn\\
\eea
The required duration  $N_{\rm RD}$ of radiation domination and
the reheating temperature $T_{\rm re}$ can then be obtained. We
clarify that in \refeq{reheating-efolds} we compute the
required value of $r=-8\alpha(n_s-1)/(2+\alpha)$ for given
$\alpha$. However, the results are essentially unchanged
if we simply set $r \simeq 0.2$.

It is worth noting that \refeq{reheating-efolds} has only
logarithmic dependence on $\epsilon_0$, $g_{\rm re}$, and
$g_{s,\rm re}$, so it suffices to take fiducial values
$\epsilon_0=1$ and $g_{\rm re}=g_{s,\rm re}=100$. The expression
is not affected by the precise values of $r$ and $A_s$, as the
dependence on these quantities is only
logarithmic. Nevertheless, the expression depends linearly on
$n_s-1$ through $N_k$, and is sensitive to $w_{\rm re}$.

\begin{figure*}[ht!]
\centering
\hspace{-0.5cm}
\includegraphics[scale=1]{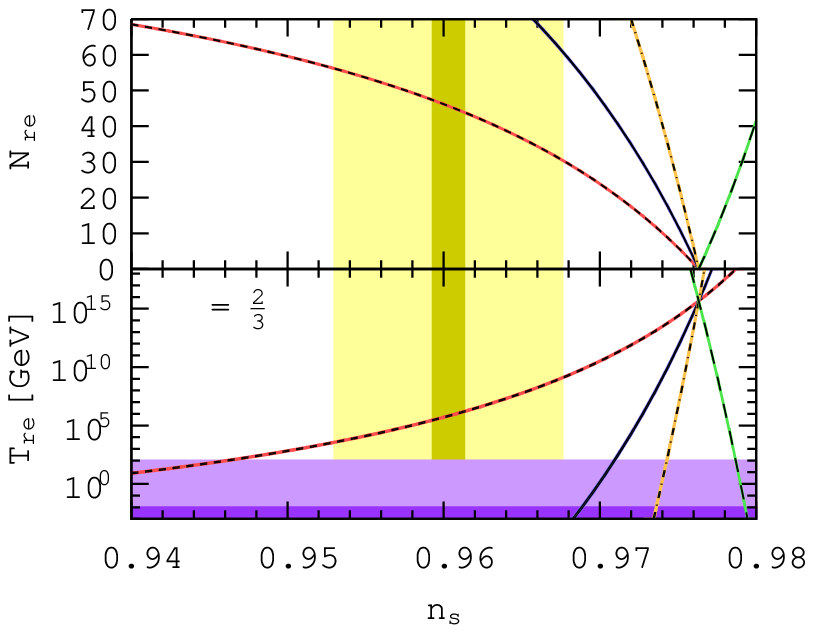}
\hspace{0.5cm}
\includegraphics[scale=1]{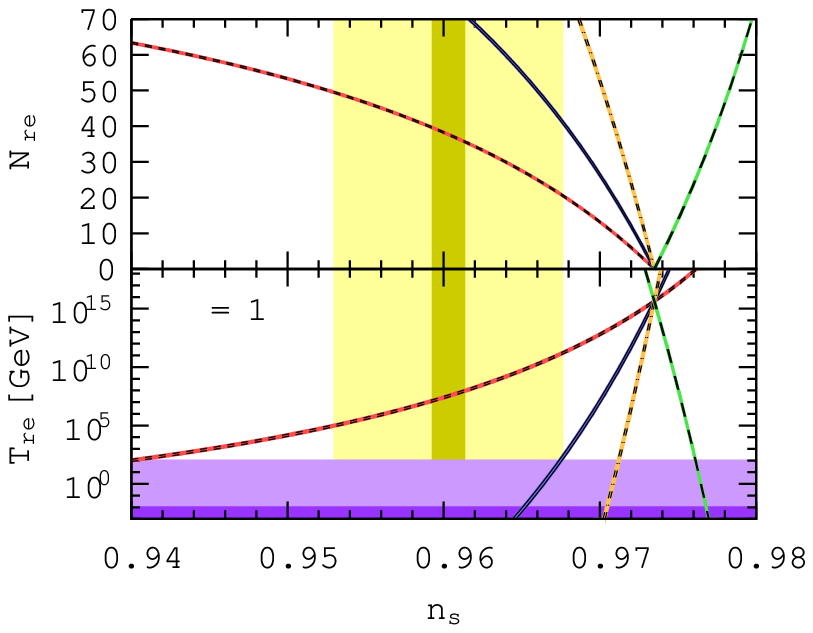} \\
\hspace{-0.5cm}
\includegraphics[scale=1]{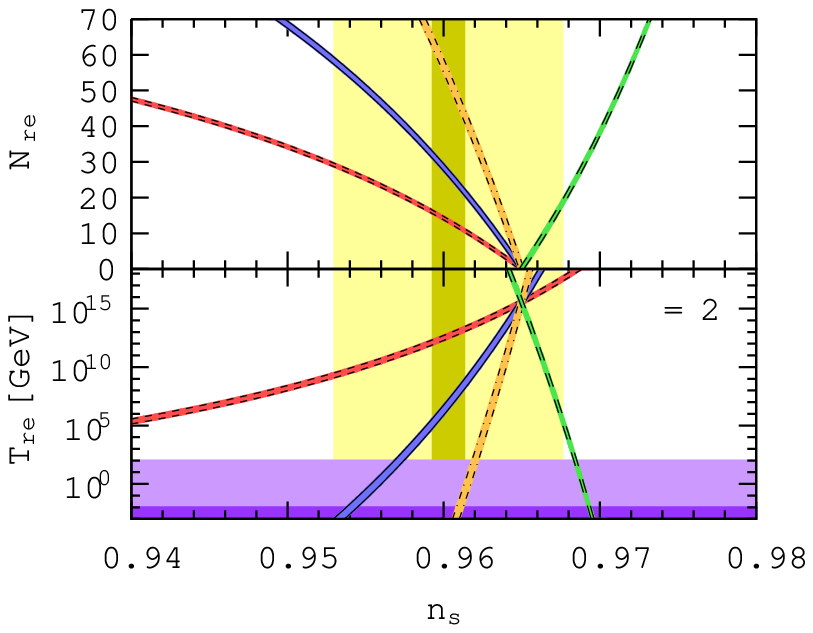}
\hspace{0.5cm}
\includegraphics[scale=1]{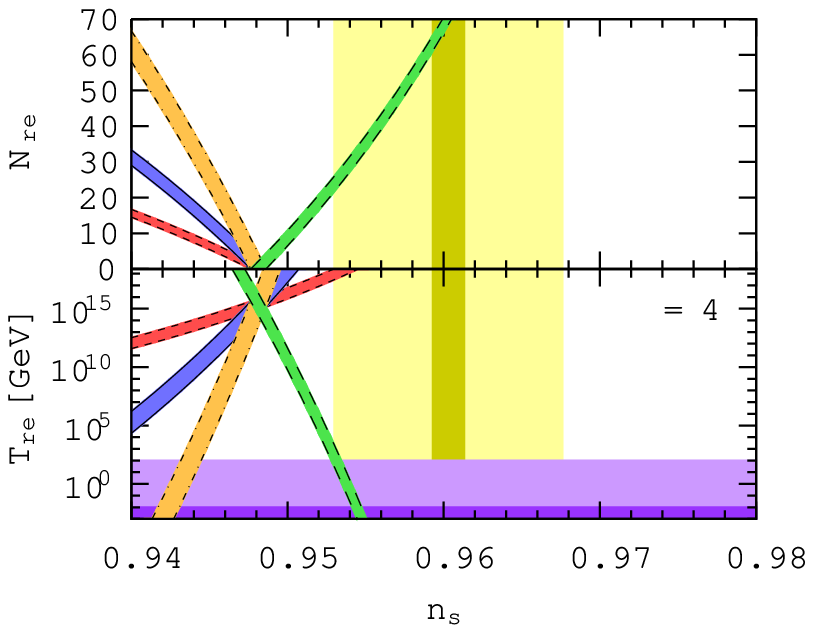}
\caption{We plot $N_{\rm re}$ (upper panels) and
     $T_{\rm re}$ (lower panels) as determined from
     \refeq{reheating-efolds} and \refeq{reheating-efoldsone},
     respectively. Results for
     power-law indexes $\alpha=2/3,1,2,4$ are each shown
     separately. Different effective equation-of-state
     parameters for reheating are considered in each case:
     $w_{\rm re}=-1/3$ (red dashed), $w_{\rm
     re}=0$ (blue solid), $w_{\rm re}=1/6$ (orange dash-dotted),
     and $w_{\rm re}=2/3$ (green long-dashed). All curves
     intersect at the point where reheating occurs
     instantaneously. The width of each curve corresponds to a
     variation of the termination condition $0.1 \lesssim
     \epsilon_0 \lesssim 1$ and also roughly the uncertainty in
     $r$. The light purple regions are below
     the electroweak scale $T_{\rm EW}\sim 100~{\rm GeV}$. The
     dark purple regions, below $10~{\rm MeV}$, would ruin the
     predictions of big bang nucleosynthesis (BBN). Temperatures 
     above the intersection point are unphysical as they
     correspond to $N_{\rm re}<0$. The light
     yellow band indicates the $1\sigma$ range $n_s-1=-0.0397\pm
     0.0073$ from Planck~\cite{Ade:2013uln}, and the dark yellow
     band assumes a projected uncertainty of
     $10^{-3}$~\cite{Creminelli:2014oaa} for $n_s-1$ as expected
     from future experiments (assuming the central value remains unchanged).}
\label{fig:Nre-and-Tre-plots}
\end{figure*}

\noindent
{\it Numerical results.}\, In \reffig{Nre-and-Tre-plots}, 
we apply the results above to compute $N_{\rm re}$ and $T_{\rm re}$ 
as functions of $n_s-1$. We study potentials with
power-law indexes $\alpha=2/3,1,2,4$. Moreover, we focus on
effective reheating equation-of-state parameters $w_{\rm re}
\geq -1/3$ (as required if inflation has ended).  As discussed
above, a matter-like $w_{\rm re}=0$ is favored for canonical
reheating, but $w_{\rm re}>1/3$ is disfavored from
model building.  Still, for illustration, we will show results
even for $w>1/3$.

Our results indicate that the quadratic model $\alpha=2$ implies
a prolonged reheating epoch for the central value $n_s \simeq
0.96$ and canonical reheating ($w_{\rm re}=0$).  A number
$N_{\rm re} \simeq 30$ of $e$-folds is required in this case,
and $T_{\rm re} \simeq 10^6~{\rm GeV}$.  A scalar 
tilt bluer than that, though, requires smaller $N_{\rm re}$ and
allows for higher reheating temperature.  For $m^2\phi^2$
inflation and canonical reheating, we approximate the numerical
results by a relation $\log_{10}\left(T_{\rm re}/10^6\,{\rm GeV}
\right) \simeq 2000\,(n_s-0.96)$ between the reheat temperature
$T_{\rm re}$ and the scalar spectral index $n_s$.  If a reheat
temperature considerably above the electroweak scale is
desired, then $n_s$ will have to be larger than its central
value.  For example, if reheating was nearly instantaneous and
set $T_{\rm re}\simeq 10^{16}$ GeV, as may be required by
GUT-scale baryogenesis models, then $m^2\phi^2$ inflation with
canonical reheating requires $n_s\simeq0.965$.  (Note here that
this $n_s$ corresponds to the pivot scale $k=0.05~{\rm
Mpc}^{-1}$ used by Planck.  The value inferred for $n_s$
increases to roughly $n_s\simeq 0.967$ for the WMAP pivot scale
 $k=0.002~{\rm Mpc}^{-1}$.)

For models with smaller power-law indexes
(e.g. $\alpha=2/3,\,1$), canonical  reheating is too efficient
in diluting the energy density if $n_s$ falls within its
$1\sigma$ error range. A reheat temperature above even the BBN
temperature requires $w_{\rm re}<0$.  Thus, unless
$n_s$ turns out to be above the current $1\sigma$ upper limit,
axion-monodromy models require some exotic mechanism of
reheating, beyond that in the canonical scenario.
On the other hand, models with larger power-law indexes
(e.g. $\alpha=3,\,4$) require $w_{\rm re}>1/3$ (dilution of
energy density faster than that that occurs with the
radiation-dominated phase) and thus also pose a challenge for
reheating models, unless $n_s$ is near the lower limit of the
current $2\sigma$ range.  Our results also indicate that
instantaneous reheating is disfavored by current measurements
except for $\alpha=2 \sim 3$. Together, these arguments (and the results
shown in \reffig{Nre-and-Tre-plots}) tend to favor the
simplest $m^2\phi^2$ models over other power-law models.

Recently, Ref.~\cite{Creminelli:2014oaa} proposed that future
measurements of $n_s-1$ and $r$ with high precision will serve
as a non-trivial consistency check of the potential shape. Their
method of determining the power-law index $\alpha$ does not rely on
good knowledge of the inflationary $e$-folds $N_k$, and is
independent of the reheating physics. Here our test of
the potential shape is complementary to theirs in the sense that
it only requires precise determination of $n_s-1$, and not of
$r$.  

\noindent
{\it Conclusions.} The recent BICEP2 measurement of a large
tensor-to-scalar ratio $r$ hints, if confirmed, at large-field power-law
inflaton potentials.  By matching the end of the inflationary
epoch to the beginning of the radiation-dominated phase we can,
with improving measurement of the scalar tilt, begin to make quantitative
inferences about the physics of reheating.  Our analysis suggests
that of the power-law inflationary models, those with
$\alpha\sim 2$, which includes the $m^2\phi^2$ model, are most
compatible with the simplest canonical reheating
scenario.  Axion-monodromy models (with power-law indexes
$\alpha=1$ or $\alpha=2/3$) require something more exotic
in the way of reheating physics, unless $n_s$ falls above its
current $1\sigma$ range.  Models with $\alpha=4$, on the
other hand are also disfavored for the $1\sigma$ range
for $n_s$.  While the statistical significance is not yet
conclusive, it is intriguing that the current data do seem to
favor a simple quadratic inflaton potential if a simple
reheating scenario is assumed.  Future more precise measurements
of $n_s$ should help make these arguments sharper.  

Although we have focused on power-law potentials, the test we
propose can in principle be applied to other potentials,
provided that $r \simeq 0.2$ already fixes the energy density
during slow-roll.

We have presented a definitive relation between $T_{\rm re}$ and
$n_s$, if inflation does indeed occur via a 
quadratic potential and is then followed by canonical reheating.
Similar relations for $w_{\rm re} \neq 0$ can be read off 
from \reffig{Nre-and-Tre-plots}.
If, moreover, the reheat temperature is considerably
above the electroweak scale, then the central value of $n_s$
should, with more precise measurements, veer upward in value,
close to $n_s=0.965$ as the reheat temperature approaches the
GUT scale. Fortunately, a precision of $\sim10^{-3}$ in the
value of $n_s$ should eventually be achieved with future
experiments such as EUCLID~\cite{Amendola:2012ys} and
PRISM~\cite{Andre:2013afa}, and with cosmic 21-cm
surveys~\cite{Mao:2008ug,Kosowsky:Physics.3.103}. 
In case high precision in $n_s$ cannot be achieved soon,
one can instead use an $r$ measured to a similar level of precision
for the same test.

Finally, laser interferometry experiments \cite{BBO} are proposed to detect
the inflationary gravitational-wave spectrum on solar-system
scales, some $40$ $e$-folds below the CMB scales \cite{GWs}. These
gravitational waves re-enter the horizon during reheating
if $T_{\rm re}<10^4~{\rm GeV}$ and will thus also probe the
physics of reheating \cite{GWreheating}.

\smallskip

The authors thank Nima Arkani-Hamed, Arthur Kosowsky, David
Kaplan, and Jared Kaplan for useful discussions, and Donghui
Jeong, Marko Simonovic, and Julian Munoz for useful comments on
an earlier draft. This work was supported by the John Templeton
Foundation and NSF grants PHY-1316665 and PHY-1214000.

\end{document}